\documentclass[prl,preprint,showpacs,superscriptaddress]{revtex4}
\usepackage{graphicx}
\usepackage{dcolumn}  

\def\gsim{\:\raisebox{-0.5ex}{$\stackrel{\textstyle>}{\sim}$}\:}
\def\Itens{\mbox{\sffamily\bfseries I}}
\def\beq{\begin{equation}}
\def\bea{\begin{eqnarray}}
\def\eeq{\end{equation}}
\def\eea{\end{eqnarray}}
\def\toner{\cite{tonertuprl1,tonertuprl2,tonertu}}
\begin{document}

\title{Hydrodynamic fluctuations and instabilities in ordered suspensions 
of self-propelled particles}
\author{R. Aditi Simha}
\email{aditi@physics.iisc.ernet.in}

\author{ Sriram Ramaswamy}
\email{sriram@physics.iisc.ernet.in}
 
\affiliation{Centre for Condensed-Matter Theory, Department of Physics, 
Indian Institute of Science, Bangalore 560 012, India.}
\date{\today}

\begin{abstract}
We construct the hydrodynamic equations for 
{\em suspensions} of self-propelled particles (SPPs) with spontaneous 
orientational order, and make a number 
of striking, testable predictions: (i) 
SPP suspensions with the symmetry of a true {\em nematic} are {\em always} 
absolutely unstable at long wavelengths.  
(ii) SPP suspensions with {\em polar}, i.e., head-tail {\em asymmetric}, 
order support novel propagating modes at long wavelengths, 
coupling orientation, flow, and concentration. (iii) In a  
wavenumber regime accessible only in low Reynolds number 
systems such as bacteria, polar-ordered suspensions 
are invariably convectively unstable. 
(iv) The variance  
in the number $N$ of particles, divided by the mean $\langle N \rangle$, 
diverges as $\langle N \rangle^{2/3}$ in polar-ordered SPP suspensions.  

\end{abstract}
\maketitle

Fish, birds, and swimming cells \cite{lighthill,pedley,xlwu} are frequently found 
to move coherently in large groups \cite{flocks,vicsek,tonertuprl1,tonertuprl2,tonertu} 
through the fluid medium they inhabit. Can general principles, like those so used so
successfully for ordered phases at equilibrium \cite{mpp}, reveal the laws governing
the long-wavelength dynamics and fluctuations of these striking and ubiquitous examples of 
liquid-crystalline \cite{maybextal} order in suspensions of self-propelled particles (SPPs)?   
Although we know of no physics experiments on {\em ordered} SPP {\em suspensions}, Gruler 
\cite{gruler} has studied ordered phases of living cells {\em on a solid substrate}, and 
finds ``living liquid-crystalline'' \cite{grulerliving} phases corresponding to two 
distinct types of cells: {\em apolar}, that is, elongated but head-tail 
symmetric, and {\em polar}, distinguishing front from rear. The ordered phases formed by 
polar SPPs have a nonzero macroscopic drift velocity $v_0$. Those formed 
by apolar SPPs have a macroscopic {\em axis} $\hat{\bf n}$ of orientation
but, like true nematics \cite{degp}, do not distinguish 
$\hat{\bf n}$ from $-\hat{\bf n}$; they thus have $v_0 = 0$. Migratory
cells such as white blood cells are polar in this sense, while melanocytes, which
distribute pigment in the skin, are apolar \cite{gruler}. Although an isolated melanocyte is 
incapable of directed motion, we still term it an SPP because it displays spontaneous, 
self-generated energy-dissipating activity, in the form of a pair of symmetrically 
pulsating dendrites \cite{gruler}. 
The pioneering studies of \cite{vicsek} and \toner \,\,consider ordered 
flocks drifting through a passive frictional background. This is appropriate for polar SPPs 
on a substrate as in \cite{gruler}, but not for SPPs immersed in a bulk fluid. 
In this Letter we use symmetry and conservation laws to construct the 
complete equations of motion for small, long-wavelength disturbances 
in polar as well as apolar ordered SPP suspensions, {\em including} the flow of the ambient fluid. 

Our main results, which we now summarize, contain striking, experimentally testable 
signatures of the nonequilibrium nature of ordered SPP suspensions, and of the crucial 
role of the hydrodynamics of the ambient fluid medium:  
{\em Purely nematic} order in SPP {\em suspensions} is {\em always destabilized}  
at small enough wavenumber $q$, by a coupled splay of the axis of orientation and a 
corresponding Taylor-Couette-like circulation of the velocity field,  
oriented near $45^{\circ}$ to the nematic axis, with a growth rate linear in $q$
\cite{nemsubstable}. 
\begin{figure}
\includegraphics{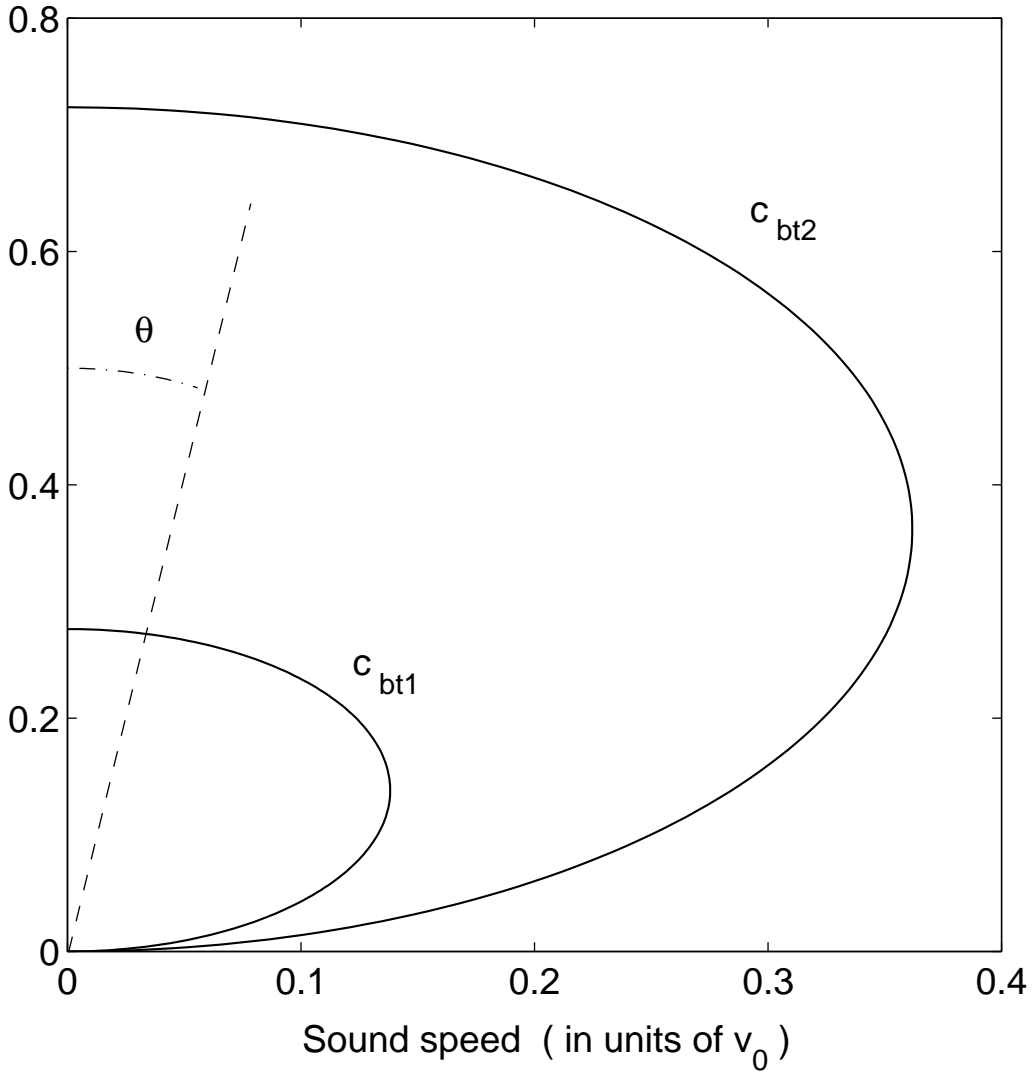}
\caption{\label{bendtwist} A polar plot of speed of the bend-twist modes as 
a function of the angle $\theta$ between the propagation direction and the
$z$ axis. }           
\end{figure}

\begin{figure}
\includegraphics{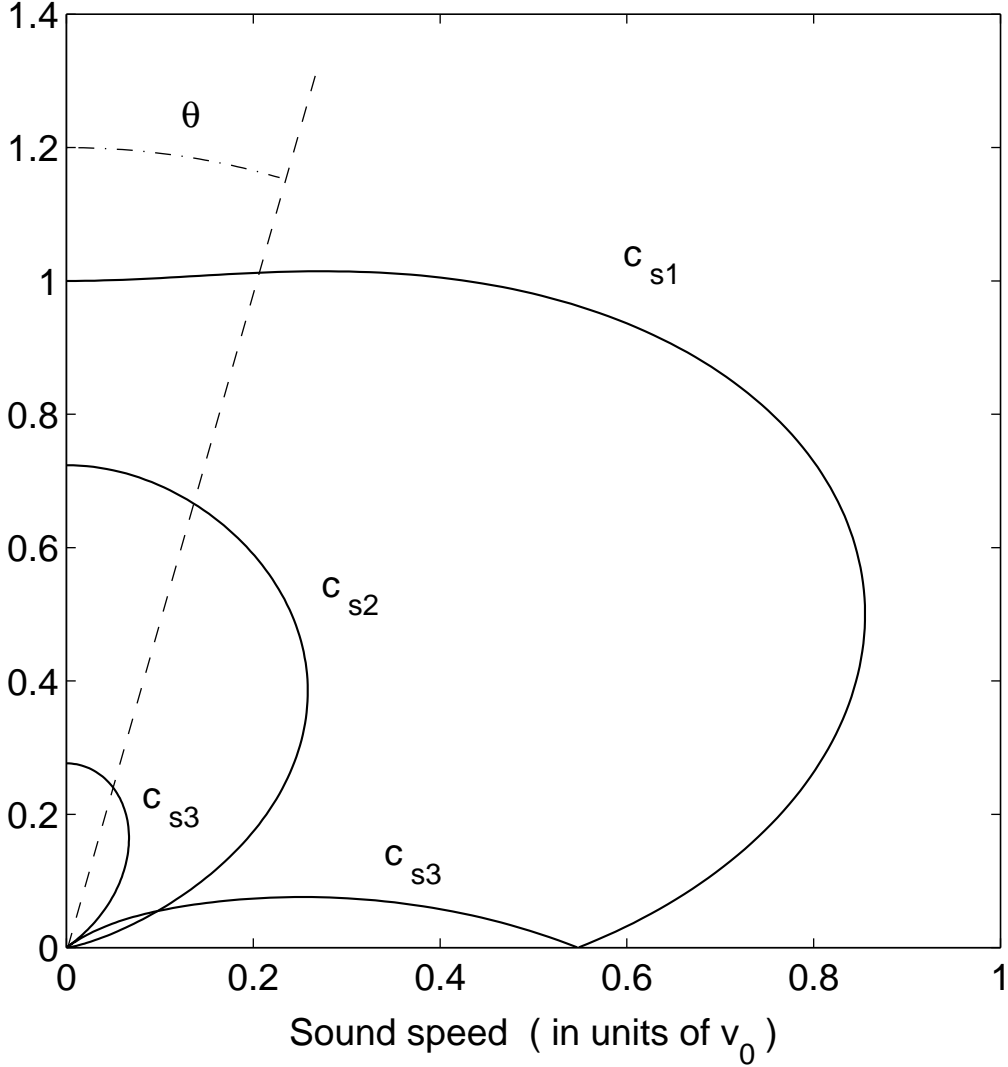}
%\end{center}
\caption{\label{splaydrift} A polar plot of the speed of the
splay-concentration and drift waves as a function of the angle $\theta$ 
between the propagation direction and the $z$ axis.}                        
\end{figure}

\noindent{\em Polar ordered} suspensions display 
some rather original propagating modes as a result of the interplay 
of hydrodynamic flow with fluctuations in the ordering direction and 
the concentration:   
a pair of bend-twist waves (with no analogue in the work of 
\cite{tonertu}), with wavespeeds as in see Fig. \ref{bendtwist} and eqn.
(\ref{cbendtwist}),  
and three waves which are a combination of splay, concentration and drift 
(a generalization of those in \cite{tonertu}), whose speeds are better understood from 
Fig. \ref{splaydrift} than from an equation.  

The results in Figs. \ref{bendtwist} and \ref{splaydrift} and Eqn. 
(\ref{cbendtwist})  
are obtained when viscous damping is neglected, which is 
valid for wavenumbers $q \ll v_0/\nu$ where $\nu$ is a  
typical kinematic viscosity. Experiments on bacterial suspensions 
are, however, likely to be in the 
Re $\ll qa \ll 1\,$ regime, where Re $=v_0 a/\nu$ is the Reynolds 
number of an SPP of size $a$. 
In that (Stokesian) limit    
we find, remarkably, that a polar-ordered suspension is always 
unstable for wavevectors ${\bf q}$ near $45^{\circ}$ to the nematic axis, with a 
growth rate $\sim v_0/a$, independent of $|{\bf q}|$. The instability is
``convective'': it travels with a speed $\sim v_0$ as it grows.
Lastly, number fluctuations in polar-ordered SPP suspensions 
are anomalously large.  The variance $\langle (\delta N)^2 \rangle$ 
in the number of particles, scaled by the mean $N$, is predicted 
to diverge as $N^{2/3}$ 
 
The results above follow from a set of hydrodynamic equations of motion,
which we now construct by generalizing \cite{mpp}. 
Assume we have nematic or polar ordered phases of SPPs, with orientation given 
by a unit  director field $\hat{\bf n}$ aligned on average in the $z$ 
direction. The nematic is invariant under $\hat{\bf n} \to - \hat{\bf n}$; the 
polar-ordered state is not. 
The slow variables \cite{mpp} are (a) the ``conserved modes'' which, for an 
incompressible suspension are \cite{zpg} the fluctuations $\delta c({\bf r},t)$ at 
point ${\bf r}$ and time $t$  in 
the local concentration $c$ of suspended particles 
about its mean $c_0$ and the total
(solute $+$ solvent) momentum density ${\bf g}({\bf r},t) \equiv \rho 
{\bf u}({\bf r},t)$, where $\rho$ is the constant mass density of the 
suspension and ${\bf u}$ the hydrodynamic velocity field; and  
(b) the ``broken-symmetry'' variables which, for both polar and nematic phases, are 
simply the deviations ${\bf \delta n}_{\perp} = \hat{\bf n} - \hat{\bf z}$. 
In polar-ordered suspensions, the drift velocity of SPPs {\em relative to the fluid} is $v_0 \hat{\bf n}$.  
Since these are driven systems, the forces entering the equations of motion need not
arise from a free-energy functional, and {\em any} term {\em not explicitly forbidden} 
by symmetry or conservation laws must be included. Moreover, there are {\it a priori} 
no relations amongst the phenomenological parameters 
other than those required by the geometrical symmetries of the problem. Note
that our
coarse-grained description applies on length-scales $\gg a$, the particle size.  

We now present the equations of motion which follow from these arguments, and
explain the physical origin of the terms therein. We begin with the polar-ordered case. 
The equations for the nematic case follow by dropping terms which violate 
$z \to -z$ symmetry.    
%, 
%{\bf The polar-ordered phase of SPPs}: 
The director field obeys 
\bea
&&\partial_t \delta {\bf n_\perp} =
- \lambda_1 v_0 \partial_z \delta {\bf n_\perp} 
- \sigma_1 {\mathbf \nabla}_\perp \delta c + 
\nonumber \\ 
&& \frac{1}{2}(\partial_z {\bf u}_{\perp}- {\mathbf \nabla}_\perp u_z) 
+ \frac{1}{2}\gamma_2 (\partial_z {\bf u}_{\perp} + {\mathbf \nabla}_\perp u_z) +
 ... 
\label{vecopeomlow}
\eea
The first term on the right-hand side of (\ref{vecopeomlow}) 
represents advection of distortions by the mean drift $v_0$ \cite{nogalinv}. 
The second says simply that SPPs move from high to low concentration -- a nonequilibrium 
``osmotic pressure''. These terms are a consequence of the lack of $z \to -z$ symmetry, 
and are present in the model of \toner as well. 
The third term on the right says that in the absence of other forces the director
rotates at the same rate as the fluid. Together with the fourth term, containing the 
phenomenological parameter $\gamma_2$, this gives rise to the ``flow-alignment'' 
\cite{degp,forsterprl} phenomenon, well-known in nematic hydrodynamics. These two terms 
are of course absent in \toner \, and are central to our work. The
ellipsis in (\ref{vecopeomlow}) subsumes diffusive terms of higher order in gradients, as well as nonlinear
terms.   

The equation of motion for ${\bf g}$ 
has the form $\partial_t g_i = - \nabla_j \sigma_{ij}$, as in any momentum-conserving 
system, where $\sigma_{ij}$ is the stress tensor.   
What distinguishes SPP systems is a contribution 
$\sigma^{(p)}_{ij} \propto n_i n_j - (1/3) \delta_{ij}$ to the shear stress, 
as a result of the self-propelling activity. We must allow for such a term because no
symmetry can rule it out in these driven systems \cite{equilpb}. 
To {\em derive} $\sigma^{(p)}_{ij}$ from the dynamics of individual SPPs  
note that, by Newton's Third Law, the forces exerted by an SPP on the fluid and by the 
fluid on an SPP are equal and opposite, i.e., the combined system of SPPs $+$ fluid is
force-free. Thus, in the equation of motion for the total momentum
density ${\bf g}$, the force density associated with each SPP and the ambient 
fluid must integrate to zero. Averaging over the detailed swimming movements, it
follows that the simplest model for, say, the $\alpha$th SPP is a rod with axis 
$\hat{\bf n}_{\alpha}(t)$, at time $t$, 
with point forces of equal magnitude $f$ on 
its ends, directed along $\pm \hat{\bf n}_{\alpha}$ \cite{pumplong}. For a collection 
of SPPs with centers at ${\bf r}_{\alpha}(t)$ and ends at 
${\bf r}_{\alpha}+ a \hat{\bf n}_{\alpha}$ and 
${\bf r}_{\alpha}- a' \hat{\bf n}_{\alpha}$,  
where we have allowed for force centers asymmetrically disposed about 
the center of the SPP, 
this yields a force density (divergence of stress)  
\bea
\label{forcedens}
-\nabla \cdot \sigma^{(p)} \equiv {\bf f}^{(p)}({\bf r},t) &=& f \sum_{\alpha} 
\hat{\bf n}_{\alpha}(t) [\delta({\bf r} - {\bf r}_{\alpha}(t) - a  \hat{\bf n}_{\alpha}(t))
\nonumber \\ 
&-& \delta({\bf r} - {\bf r}_{\alpha}(t) + a' \hat{\bf n}_{\alpha}(t))]. 
\eea
The apolar and polar SPPs we mentioned early in this article 
correspond respectively to $a = a'$ and $a \neq a'$.    
Expanding the delta-functions in (\ref{forcedens}) about 
${\bf r}_{\alpha}$ gives  
\beq
\label{stress} 
f_i^{(p)} \simeq -{a + a' \over 2} f 
\nabla_j\sum_{\alpha} \hat{n}_{\alpha i}\hat{n}_{\alpha j} 
\delta({\bf r} - {\bf r}_{\alpha}) 
+ O(\nabla \nabla)  
\eeq
In a coarse-grained description this clearly leads to 
a shear stress  
\beq
\label{stressvv}
\sigma_{ij}^{(p)}({\bf r},t) = 
{a + a' \over 2} f c({\bf r},t) (n_i n_j - {1 \over 3} \delta_{ij})  
+ O(\nabla)   
\eeq
as claimed above. Note that only the {\em sum} $a + a'$ appears in 
(\ref{stressvv}); to leading order in gradients, the self-propelling stress
is the same for polar and apolar phases. 
Using this stress in the momentum equation  
$\partial_t g_i = - \nabla_j \sigma_{ij}$, 
Fourier-transforming in space, replacing ${\bf g}$ by the hydrodynamic
velocity field ${\bf u} = {\bf g} / \rho$,  
and projecting transverse to the 
wavevector ${\bf q}$ to impose incompressibility ($\nabla \cdot {\bf u} = 0$) 
yields, after some algebra,  
\beq
\label{momeomlow}
%(
%+\nu q^{2})
{\partial {\bf u}_{\perp} \over \partial t} =  
-iw_0 q_z(\Itens - 2{{\bf q}_{\perp}{\bf q}_{\perp} \over q^2}).
{\bf \delta n}_{\perp} 
- i\frac{q_z^2}{q^2} \alpha\,
({\bf q}_{\perp} \delta c) + ... 
\eeq 
where $\Itens$ is the unit tensor, and $\alpha \sim f a /\rho$ and $w_0 \sim c_0
\alpha$ are phenomenological constants proportional to the activity of the SPPs.     
In (\ref{momeomlow}), nematic elastic torques \cite{degp} and 
viscous forces, both of which are subdominant at small $q$, have been ignored, as 
have inertial and other nonlinearities.  
The pressure-like $\alpha$ term despite transverse 
projection, and the acceleration proportional to the bend  
$q_z \delta {\bf n}_\perp$, arise purely because of self-propulsion \cite{equilpb}.  
 
Number conservation of SPPs implies $\partial_t \delta c = - \nabla \cdot {\bf j}$, where 
the current ${\bf j} = c v_0 \hat{\bf n}$ apart from advection by fluid flow and
diffusion. To lowest order in gradients and fields, this tells us that  
\begin{equation}
\label{conceomlow}
\left(\partial_t +v_0 \partial_z 
\right)\,\delta c 
+ c_0 v_0 \nabla_{\perp} \cdot \delta {\bf n}_\perp =0
\end{equation}
which says the concentration changes by advection by the mean drift 
$v_0$ or by local splay (since that implies a divergence in the 
local SPP velocity).   

We can now extract the propagating waves and instabilities mentioned at the start of 
this paper, from the dispersion
relation, frequency $\omega$ as a function of wavevector ${\bf q}$, for
modes varying as exp$(i {\bf q} \cdot {\bf r} - i \omega t)$,
implied by equations (\ref{vecopeomlow}), (\ref{momeomlow}) and 
(\ref{conceomlow}). 

{\bf Bend-twist waves}: Taking the curl of (\ref{vecopeomlow}) and 
(\ref{momeomlow}) yields the coupled dynamics of bend or twist 
(${\mathbf \nabla} \times {\bf n}_{\perp}$) and vorticity 
(${\mathbf \nabla} \times {\bf u}_{\perp}$). Fourier-transforming in space 
and time yields $\omega = c_{bt}(\theta) q$, with wavespeeds 
(see Fig. \ref{bendtwist})   
\beq
\label{cbendtwist}
c_{bt}(\theta) =  (c_1 \pm c_2) \cos \theta
\eeq
where $\theta$ is the angle between the propagation vector and 
the ordering direction, and $c_1$ and $c_2$ (of order the drift speed  
$v_0$ of the SPPs) are phenomenological constants.  
An analysis ignoring hydrodynamic flow, as in \toner, 
would instead predict purely diffusive relaxation for bend and twist. 

{\bf Splay-concentration and drift waves}: Taking the {\em divergence} of 
(\ref{vecopeomlow}) and (\ref{momeomlow}) results in coupled 
equations of motion for 
${\mathbf \nabla} \cdot {\bf n}_{\perp}$,
${\mathbf \nabla} \cdot {\bf u}_{\perp}\,$, and $\delta c$. 
The resulting wavelike eigenmodes are a generalization of those seen in 
\cite{tonertu}. Their speeds obey a messy 
cubic equation which, we can show \cite{long}, has {\em real} solutions 
for a finite range of parameter values. 
The speeds as functions of direction, for typical parameter values, are illustrated in 
Fig. \ref{splaydrift}. 

{\bf Viscous effects, and the instability of polar SPP suspensions 
in the Stokesian limit}: 
So far, we have ignored damping terms in the equations 
of motion since they enter at next-to-leading order in wavenumber $q$, 
There are three types of such terms: (i) viscous damping 
$\sim \nu q^2 {\bf u}$ 
in the momentum equation (where $\nu$ is  
a typical kinematic viscosity of the suspension);  
(ii) $D q^2 \delta c$ in the 
concentration equation, $D$ being a diffusivity;   
(iii) $D_n q^2 {\bf \delta n}_{\perp}$ 
in the director equation where $D_n$ is a director diffusivity which we expect to be
comparable to $D$. In bacteria, diffusivities are of order $10^{-6}$ cm$^2$/s 
\cite{xlwu} and can
clearly be neglected relative to the viscosity. In large-Re systems such as fish, we
have no reliable estimate of $D$, but it seems reasonable to assume it is no larger
than $\nu$. Accordingly, when damping terms are included, the modes will remain
propagating only for $qa < v_0a/\nu \equiv$ Re, i.e., at practically all length scales in 
the case of fish, but only at exceedingly large length scales for bacteria, 
where Re $\sim 10^{-6} \, {\rm to} \, 10^{-4}$.  
The regime of primary interest for bacteria is 
$\nu q^2 \gg v_0 q$, i.e. $qa \gg v_0a/\nu \equiv$ Re, 
where viscous damping dominates. 
For such low Reynolds number systems we can use the Stokesian approximation, where 
the velocity field ${\bf u}$ is determined instantaneously 
by a balance between viscous and other (in this case self-propulsive) 
stresses. We can thus replace the acceleration 
in (\ref{momeomlow}) by $\nu q^2 {\bf u}$ and use it 
to eliminate ${\bf u}$ from (\ref{vecopeomlow}) and (\ref{conceomlow}).  
This yields effective equations of motion for the splay $\nabla \cdot 
{\bf \delta n}_{\perp}$ 
and $\delta c$, the coupled dynamics of which can be seen to  
yield an unstable mode with growth rate 
\beq
\label{gammastokes}
\Gamma(\theta) \sim B 
(\gamma_2 \cos 2\theta +1)\cos 2 \theta 
\eeq
with $|B| \sim f \phi /a^2 \eta \sim v_0\phi/a$, $\phi$ being the volume fraction of SPPs.  
This implies an instability just above or just below $\theta = \pi/4$,  
depending on the sign of $B$.  
The mode has a nonzero real part $\sim \pm v_0 q \cos \theta$: 
this is a {\em convective} instability, which is seen if one {\em follows} the
travelling waves.  
Note that $\Gamma$ is independent of the {\em magnitude} of ${\bf q}$, 
as a result of the long-ranged hydrodynamic interaction in the Stokesian limit. 
For $qa \gsim \phi^{1/2}$, director and particle diffusion enter 
to restabilize the mode, so the unstable regime is Re $\ll qa \ll 
\phi ^{1/2}$.   

{\bf Instability of SPP nematic suspensions}: 
The dynamics of an SPP {\em nematic} suspension can be obtained from 
(\ref{vecopeomlow}), (\ref{momeomlow}) and (\ref{conceomlow}) 
simply by omitting all terms not invariant 
under $z \to -z$, i.e., by setting $v_0 = \sigma_1 = \alpha = 0$. 
%This can readily be shown to yield an mode with 
%$\omega \sim -iA q\cos 2 \theta(\gamma_2 \cos 2\theta +1)$, 
%linear in the wavenumber, where $|A| \sim \sqrt{w_0}$ .
This can readily be shown to yield the dispersion,  
$\omega ^2 \sim -A q^2 \cos 2 \theta (\gamma_2 \cos 2\theta +1)$, 
where $|A| \sim w_0$ .  
Physically, this can be understood 
as follows. The nematic symmetry $z \to -z$ 
means that a distortion with wavevector at exactly $45^{\circ}$ has no 
wavelike response: it doesn't know which way to go, so the $O(q^2)$  
contribution to the squared frequency $\omega^2({\bf q})$ vanishes 
at this angle. $\omega^2$ must thus {\em change sign} as 
$\theta$ crosses
$45^{\circ}$.  For $\theta = \pi / 4 - \epsilon$, where $\epsilon$ is 
small and has the same sign as $A$, we thus get an instability 
with $\omega({\bf q}) \sim i \sqrt{|\epsilon|}\, q - i q^2$. Here the 
last term appears when diffusive and viscous terms and the 
conventional nematic-elastic torques are included, and will stabilize modes
at larger wavenumber. 
For an SPP nematic {\em on a substrate}
as in \cite{gruler}, a wavenumber-independent damping enters the momentum 
equation (\ref{momeomlow}) and in general suppresses this instability 
\cite{long}.
 
{\bf Fluctuations}: In striking contrast to systems at thermal 
equilibrium, ordered suspensions of SPPs display giant number 
fluctuations. To see this, we must include noise, as well as
the damping terms whose form we have discussed, in the equations of 
motion. For a general SPP suspension one expects noise sources 
of three kinds:  
thermal Brownian motion (negligible for particles larger 
than a few microns), many-particle hydrodynamic interactions, 
and intrinsic fluctuations in the self-propelling activity 
of individual SPPs.  
In our coarse-grained description it is reasonable to assume 
Gaussian noise sources ${\bf f}_u$, $f_c$, and ${\bf f}_n$, 
delta-correlated in time, in the equations of motion for ${\bf u}$, 
$c$, and ${\bf n}$. Momentum and number conservation imply 
variances $\sim q^2$ for ${\bf f}_u$ and $f_c$ for wavenumber 
$q \to 0$, and nonzero for ${\bf f}_n$. With these noise and damping terms 
in place we can calculate correlation functions of the various 
fields in the steady state. We find that the structure 
factor 
$S({\bf q}) = c_0^{-1}\int_{\bf r}\exp -i {\bf q} \cdot {\bf r} 
\langle \delta c({\bf 0}) \delta c({\bf r})\rangle \sim 1/q^2$, 
which implies that the variance of the number of particles, 
scaled by the mean, grows as $L^2 \sim N^{2/3}$ 
for a three-dimensional region of linear size $L$ containing 
$N$ particles on average. 
Physically, this 
happens because distortions in the director produce mass flow. 
Since the director fluctuations -- a broken-symmetry mode -- 
are large, this results in giant number fluctuations as well.  
Similar supernormal number variances were predicted \toner \, for SPPs on a 
substrate; it is remarkable that they survive here in the presence of 
long-ranged hydrodynamic interactions.   

Let us now estimate the numerical values of the effects we predict. 
All wavespeeds are expected to be of order the drift speed of an 
SPP (from $\mu$m/s to cm/s as one goes from bacteria to fish). 
If we take the kinematic viscosity $\nu$ to be that of water, 
then the growth rate of the convectively unstable mode in 
Stokesian polar-ordered SPP suspensions, say for bacteria
(where velocities are $\,\sim 10 \mu$m/s, size $a \sim \mu$m, and 
$\phi\,$ a few percent), is $0.1$s$^{-1}$.  
We look forward to experimental tests of our predictions, on living 
organisms or perhaps on artificial SPP suspensions.  

An analysis of the complete (not linearized) 
equations of motion, now underway, will show 
how nonlinear fluctuation effects 
renormalize the speeds and dampings of the modes presented 
here, and whether they alter the manner in which number 
fluctuations diverge with size. The full equations of motion 
will also prove useful in understanding the effect of 
imposed shear flows on the state of order of SPP suspensions, 
and the role of boundaries and topological defects in these systems. 
These as well as other results on SPP nematics on substrates, and an
analysis of isotropic SPP suspensions \cite{xlwu},  
will appear elsewhere \cite{long}. 
 
SR thanks the Aspen Center for Physics for support while 
part of this work was done.


\begin{references}
\bibitem{maybextal} Ref. \cite{tonertu} suggests that crystalline, translational order
may be seen in flocks as well.
\bibitem{zpg} We ignore heat diffusion, nutrient uptake, 
excretion, reproduction and death.
\bibitem{nogalinv} As in \toner, the speed of advection is not in
general equal to the drift speed, hence the factor $\lambda_1$.
\end{references}
\end{document}